\documentclass{sig-alt-release2}
\usepackage{epsfig}

\begin{document}

\conferenceinfo{CIKM'10,} {October 25--29, 2010, Toronto, Ontario, Canada.} 
\CopyrightYear{2010}
\crdata{978-1-4503-0099-5/10/10}
\clubpenalty=10000
\widowpenalty = 10000

\title{LiquidXML: Adaptive XML Content Redistribution}

\author{Jes\'us~Camacho-Rodr\'iguez$^1$ $~~$ Asterios~Katsifodimos$^{1,2}$ $~~$ Ioana~Manolescu$^{1,2}$ $~~$ Alexandra Roatis$^{1,3}$\\
\affaddr{$^1$INRIA Saclay, France $~~^2$LRI, Universite Paris XI, France $~~^3$West University of Timisoara (UVT), Romania}\\
\email{firstname.lastname@inria.fr}
}

\maketitle

\vspace{-2.5mm}
\begin{abstract} We propose to demonstrate LiquidXML, a platform for managing large corpora of XML documents in large-scale P2P networks. All LiquidXML peers may publish XML documents to be shared with all the network peers. The challenge then is to efficiently (re-)distribute the published content in the network, possibly in overlapping, redundant fragments, to support efficient processing of queries at each peer. The novelty of LiquidXML relies in its {\em adaptive} method of choosing which data fragments are stored where, to improve performance. The ``liquid'' aspect of XML management is twofold: XML data flows from many sources towards many consumers, and its distribution in the network continuously adapts to improve query  performance.
\end{abstract}
\vspace{-2.5mm}
\vspace{-3mm}
\vspace{-3mm}

\section{Introduction}
\vspace{-2mm}
We consider the problem of building large-scale, decentralized XML stores, capable of efficiently evaluating XML queries over documents indexed in a DHT based peer-to-peer network. Our solution is based on the previously built platform ViP2P (standing for {\em Vi}ews in {\em P}eer-to-{\em P}eer) which we developed~\cite{vip2p}. In ViP2P, any peer may publish XML documents, which it is willing to share with the other peers. Moreover, any peer may establish long-running subscriptions to XML content published anywhere in the network, that matches a given subscription query. The results of such subscriptions are stored at the subscriber peer, and advertised in the DHT network, so that other peers may re-use them to answer their own queries, with less computation effort. Conceptually, thus, the result of each subscription can be seen as a {\em materialized view}, based on which subsequent queries can be rewritten. It is important to note that: ($i$)~{\em the queries defining the subscriptions} (and not the subscription results) are indexed in the DHT network, leading to a small overhead of data sharing; and ($ii$)~we consider a {\em collaborative} scenario, where each peer is willing to share its subscriptions/views with any other. ViP2P has been shown to scale on  up to 500 peers, and 100 GB of XML data. A separate development on top of ViP2P, illustrating P2P document annotations, was demonstrated~\cite{AnnoViP}.

\begin{figure}[t!]
\begin{center}
\includegraphics[width=0.9\columnwidth]{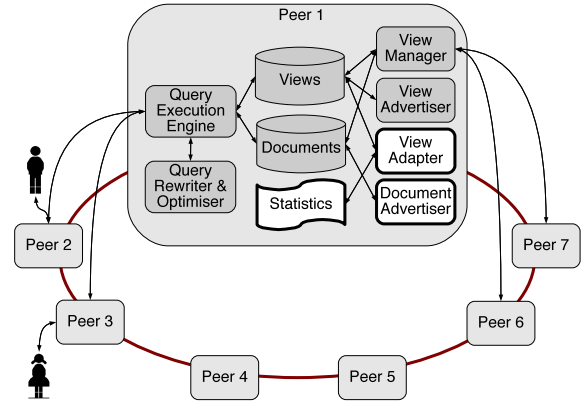}
\caption{LiquidXML platform architecture.\label{fig:archi}}
\end{center}
\vspace{-5.7mm}
\end{figure}

Our proposed demo features LiquidXML, a system built on top of ViP2P. Its main technical innovation is to {\em automatically select and continuously adapt} the set of materialized views on each peer, to improve query processing performance both for the view holding peer, and for the other network peers. LiquidXML continuously adapts by adding more materialized views and/or replacing low-utility views with more useful ones according to the query workload. Figure~\ref{fig:archi} outlines LiquidXML's architecture, on top of ViP2P. The modules shown in thick, white boxes are novel to LiquidXML and the main focus of the demo.

\vspace{-3.8mm}

\section{LiquidXML platform outline}
The main aspects of the LiquidXML content management platform can be summarized as follows.

\begin{figure*}[ht!]
\includegraphics[width=0.33\textwidth]{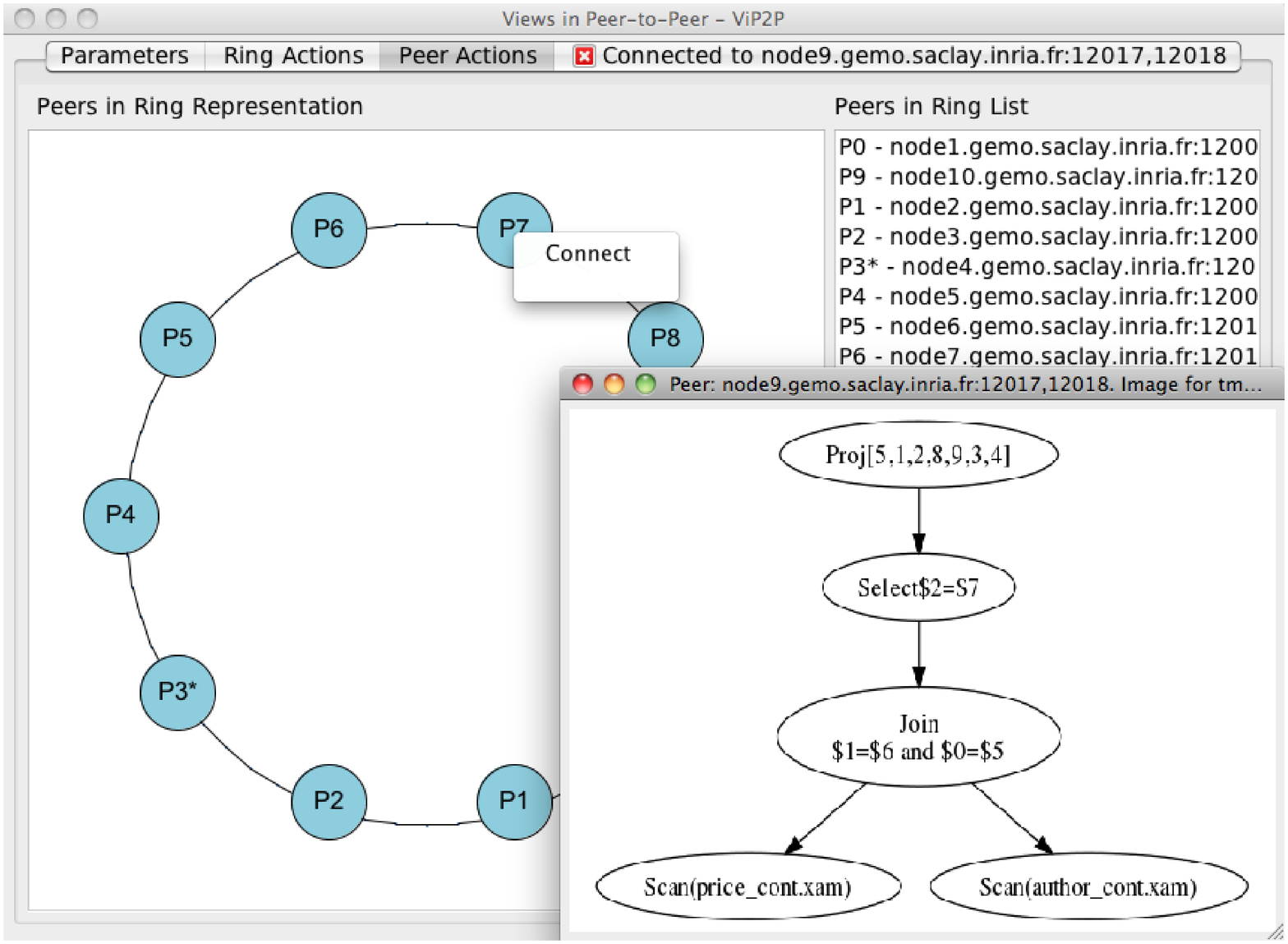} 
\includegraphics[width=0.33\textwidth]{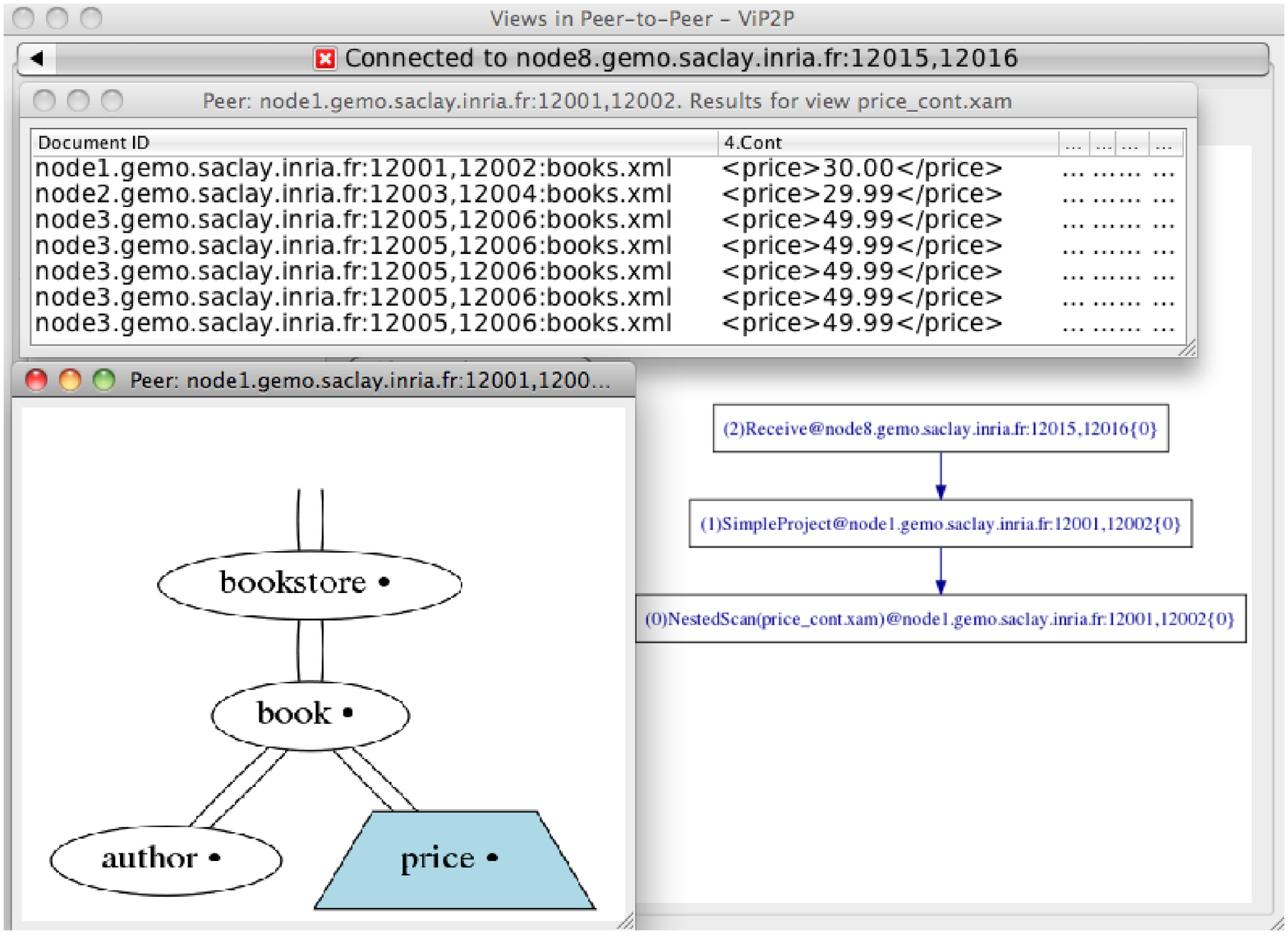}
\includegraphics[width=0.33\textwidth]{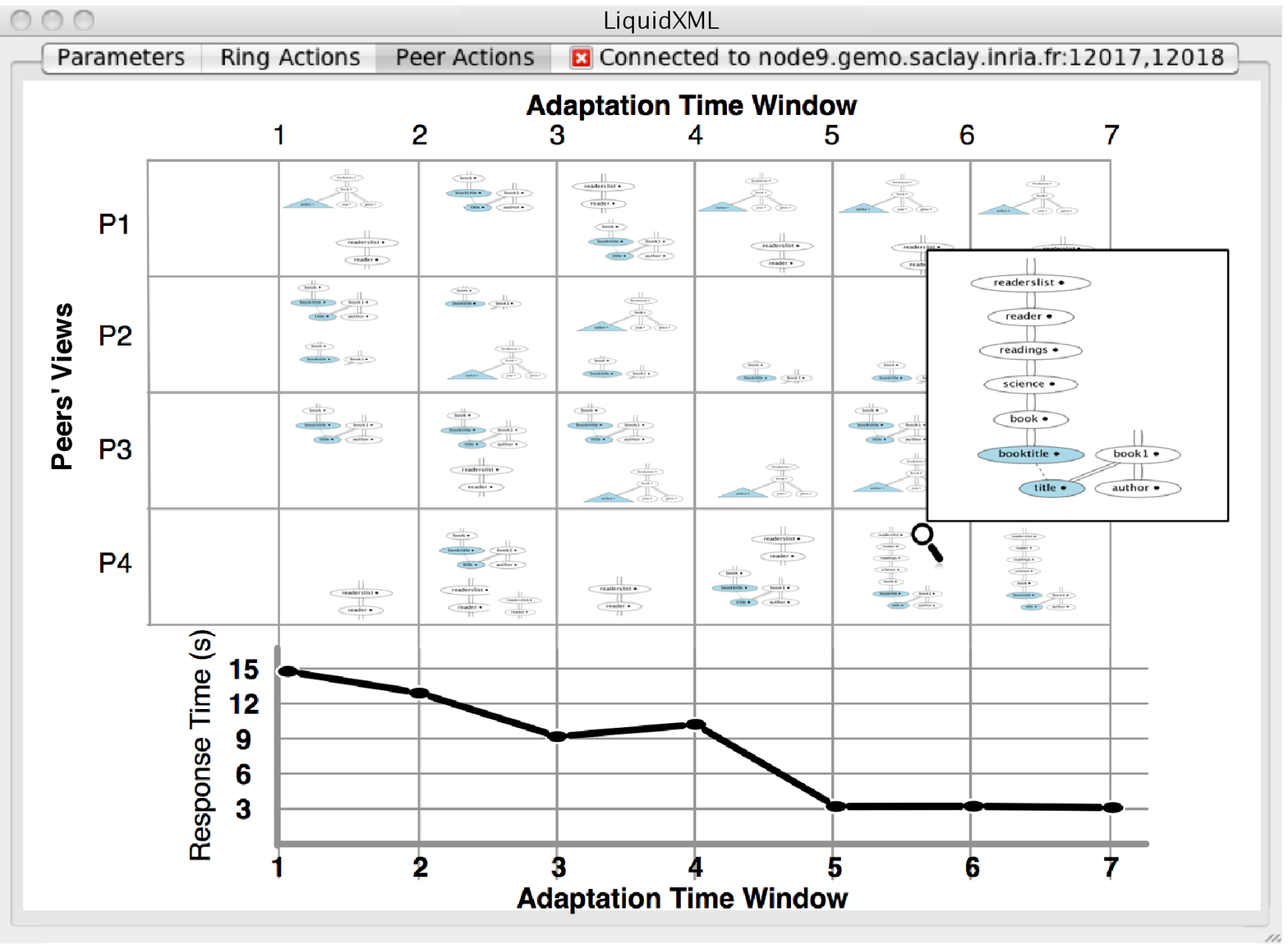}
\vspace{-5mm}
\caption{Demonstration screenshots: peer network and sample query rewriting (left); sample view data and simple physical plan (center); LiquidXML's adaptation monitoring window (right).\label{fig:screenshots}}
\vspace{-4.5mm}
\end{figure*}

\noindent \textbf{Peer space budget}~When joining the network, each peer declares a space budget that it can spend to store data structures aimed at improving query performance for itself and for the other network peers. Upon joining, the space budget of the peer is unused (empty).

\noindent \textbf{Document-level indexes}~ LiquidXML builds \emph{document-level indexes}, distributed in the network. For each term (element, attribute name or word) appearing in an XML document, the  URIs of all the network documents featuring that term is stored by some peer. This index allows to locate all documents which {\em may} contain answers to a given query. Sending the query to the corresponding peers leads to obtaining the results. This query answering mechanism is not always performant, since ($i$)~ some documents may not lead to answers, even if they contain all the query terms, and ($ii$)~queries are always evaluated from scratch. 

\noindent \textbf{Query statistics at a peer}~Each peer $p$ is aware of a query set $Q_p$, which is a subset of all the queries being asked in the network. This set contains the queries asked by $p$, and the queries which $p$ helped answering based on the data stored at $p$. Peer $p$ collects, for each query $q\in Q$,  its frequency ($\#q$) in a given time window of length $\tau$. %, namely, $p$'s documents, document-level indexes, or stored subscriptions (materialized views).

\noindent \textbf{Peer candidate views}~Based on its query statistics, each peer $p$ identifies a set of XML views to be materialized and shared with other peers that may need them, in order to reduce the response time of $p$ and other peers' queries. The general problem of finding all candidate materialized views for a given  XML query workload is very complex~\cite{DK}. In LiquidXML, we use a data cube-style~\cite{DataCubes} lattice to identify the most appropriate candidates. For each of the candidate views, $p$ computes: $(i)$~a \emph{cost} estimation (in terms of size) and $(ii)$~the {\em benefit} (in terms of network and computation savings) that the candidate view would bring to the system if it is materialized. 

\noindent \textbf{View size estimation}~For each document published by peer $p$, a compact document synopsis is also indexed in the DHT. The document synopsis is based on XSum~\cite{PathSummaries}, our own Dataguide implementation and is used to estimate the contribution (in size) of a document to a candidate view. To estimate the size of a candidate view, the synopses of all documents that may contribute to the view are retrieved. The total estimated size of a candidate view, denoted $size(v)_\epsilon$, is the sum of all document contributions to the view.

\noindent \textbf{Query cost estimation}~The presence of a new materialized view may change the way a query is processed, if the query can be rewritten based on that view. We assign to each rewriting a cost estimation, reflecting the amount of data transmitted between peers to evaluate the rewriting. Given a set of materialized views ${\cal V}$ and a query $q$, we denote the estimated cost of answering the query $q$ as $cost(q,{\cal V})_\epsilon$.

\noindent \textbf{View benefit estimation} Given a candidate view $v$, the total set ${\cal V}$ of views currently materialized in the network, and a query workload $Q$, we estimate the benefit of $v$ for $Q$ with respect to ${\cal V}$ as:

\vspace{-2mm}
\begin{center}
$b(v, Q, {\cal V})=\sum_{q\in Q}(\#q)\times(cost(q,{\cal V})_\epsilon - cost(q,{\cal V}\cup \{v\})_\epsilon)$
\vspace{-0.7mm}
\end{center}

\noindent \textbf{Putting it all together: LiquidXML adaptation} Each LiquidXML peer continuously gathers statistics and costs as outlined above. At regular $\tau$ intervals, each peer enumerates candidate views and materializes those maximizing the benefit-to-size ratio, up to the limit of its space budget. Existing views with a low benefit-to-size ratio can be dropped to make room for more interesting ones.

\vspace{-3mm}

\section{Implementation and scenario}
\label{sec:scenario}
\vspace{-1mm}

LiquidXML is implemented in Java, on top of ViP2P, using the FreePastry DHT as the underlying P2P network and the BerkeleyDB library  to store materialized views. The supported query language is a core subset of XQuery, consisting of conjunctive tree patterns with joins.

LiquidXML's GUI\footnote[1]{http://vip2p.saclay.inria.fr/?page=liquidxml} will enable demo attendants to: ($i$)~connect to any peer and inspect its views, queries, and statistics; ($ii$)~control adaptation parameters, e.g. synopsis size, the adaptation time window etc. ($iii$)~view the evolution of the peers' views over time, ($iv$)~view logical plans resulting from rewriting and the resulting distributed physical query plans. Figure~\ref{fig:screenshots} shows some sample screenshots.

We will show the demo on 250 machines of the Grid5000 network (http://www.grid5000.fr). We will trace query execution and performance in three scenarios:

\noindent \textbf{1.~Document-level indexes}:~Only document-level indexes will be used to locate the documents potentially containing query results, to which the query is shipped.
  
\noindent \textbf{2.~User-defined views}:~Users may manually define specific views to materialize. Queries will be answered by rewriting them in terms of the user-defined views.

\noindent \textbf{3.~Full adaptive LiquidXML}:~Peers automatically adjust their  views to match the needs of the distributed query workload. Demo attendees will visualize the set of views on each peer, as it varies over the time.
More information about LiquidXML can be found in our technical report\footnote[2]{http://vip2p.saclay.inria.fr/liquidxml/report.pdf}.

\vspace{1mm}
\noindent\textbf{Acknowledgements} This work has been partially funded by Agence Nationale de la
Recherche, decision ANR-08-DEFIS-004. We are grateful to S. Zoupanos, A. Tilea and V. Mishra for their contributions to the ViP2P project.

\vspace{-3mm}
{\small 
\bibliographystyle{abbrv}
\bibliography{references}
}

\end{document}